\documentclass[aps,prb,amsmath,amssymb,twocolumn,superscriptaddress,longbibliography]{revtex4-1}

\usepackage{graphicx}% Include figure files
\usepackage{dcolumn}% Align table columns on decimal point
\usepackage{bm}% bold math
\usepackage{hyperref}% add hypertext capabilities

\begin{document}

\title{Pseudogap in the lightly hole-doped triangular-lattice moir\'e Hubbard model}

\author{V. I. Kuz'min}
\email{kuz@iph.krasn.ru}
\affiliation{Kirensky Institute of Physics, Federal Research Center KSC SB RAS, Krasnoyarsk, 660036 Russia}

\author{M. A. Visotin}
\affiliation{Kirensky Institute of Physics, Federal Research Center KSC SB RAS, Krasnoyarsk, 660036 Russia}
\affiliation{Siberian Federal University, Krasnoyarsk, 660041 Russia}

\author{S. G. Ovchinnikov}
\affiliation{Kirensky Institute of Physics, Federal Research Center KSC SB RAS, Krasnoyarsk, 660036 Russia}
\affiliation{Siberian Federal University, Krasnoyarsk, 660041 Russia}

\date{\today}

\begin{abstract}
The electronic structure of the lightly hole-doped triangular-lattice moir\'e Hubbard model is studied within cluster perturbation theory (CPT) using 13-site clusters for a fixed doping concentration $p=1/13$ varying the Coulomb parameter $U$ and the hopping phase parameter $\phi$ related to the spin-orbital interaction. We have also developed a rather simple generalized mean-field approximation (GMFA) containing the amplitude of the spin correlations as a free parameter to fit the CPT data.- The evolution of the Fermi surface and the pseudogap with the parameters $\phi$ and $U$ is explained from the viewpoint of the short-range magnetic order. The geometric frustration and the additional model parameter related to the spin-orbital interaction result in a more rich physics of the pseudogap state compared to the case of a more conventional square lattice.
\end{abstract}

\maketitle

\section{\label{Intro}Introduction} 

The Mott-type metal-insulator transition and unusual electronic properties in the regime of strong electron correlations are phenomena of fundamental interest in condensed matter physics, the understanding of which is far from clear. Strong correlations make the use of single-particle methods very limited. The Hubbard model \cite{Hubbard63} is the simplest model for investigating the metal-insulator transition, which can be caused by various parameters, such as interaction, temperature, doping, or geometric frustration \cite{Kampf94,Metzner89,Georges96}.

One of the aspects of the Hubbard model is the momentum-space anisotropy of the electronic structure together with the metal-insulator transition, widely studied on a square lattice \cite{Senechal04,Sadovskii05,Gull10,Kuzmin20,Ye23}, which has attracted a lot of attention due to the long-standing problem of the pseudogap in cuprates and its relation to superconductivity \cite{Timusk99,Robinson19}. The pseudogap is an unusual momentum-selective suppression of the density of states and the spectral function at the Fermi level. It is closely related to the Fermi arc, which is the type of Fermi surface specific to strongly correlated electronic systems. 

Recently, it has been proposed that the Hubbard model on a triangular lattice may describe twisted transition-metal dichalcogenides (tTMD) \cite{Wu18}. In the triangular Hubbard moir\'e model some additional degrees of freedom appear. In the simplest model case it can be considered as a phase modulation of the hopping parameter $t_{i,j}$, which results in a new situation compared to cuprates. Until now, little has been known about the momentum dependence of the Fermi surface and the electronic structure of the doped triangular Hubbard moir\'e model. The twist angles in $\text{tWSe}_2$ were studied by different techniques, including atomic force microscopy, optically measured second harmonic generation, and transport properties. Experimental studies of low-temperature resistivity in tTMD under variation of bandwidth by a displacement field \cite{Wang20, Ghiotto21} have found strange metal behavior in proximity to the Mott transition with linear temperature-dependent resistivity similar to a strange metal in high-$T_c$ cuprates \cite{Chen22}. The signatures of the pseudogap behavior have been found in the density of states of the triangular Hubbard model, using dynamic cluster approximation (DCA) in the case of half-filling  \cite{Downey23}, as well as in the light hole doping \cite{Downey24}. The density matrix renormalization group studies \cite{Shirakawa17,Chen22} point to the existence of an intermediate state between the metallic and insulating antiferromagnetic phases. There is evidence for similarity between the conductivity phase diagrams of cuprates and $\text{tWSe}_2$ \cite{Wang20}. Thus, the existing and, probably, coming in the future moir\'e structures may provide the basis for a deeper understanding of the pseudogap phenomenon.

In the present paper, two complementary methods for strongly correlated systems are applied to the triangular moir\'e Hubbard model at low doping to study the main tendencies in the electronic structure of systems such as twisted homobilayer dichalcogenides. We study the momentum and frequency dependence of the spectral function together with short-range spin correlations within the paramagnetic phase. The main approach here is cluster perturbation theory (CPT) \cite{Senechal00,Senechal02}, which allows investigation of the electronic structure while explicitly taking into account the short-range correlations within a finite cluster. The other method we use is a version of generalized mean-field approximation (GMFA), based on the equations of motion of the Green functions and Mori projection technique \cite{Zubarev60, Mori65} that we supply with a couple of fitting parameters in order to capture the main phenomenology and to allow a better comparison with CPT.

Here, it is shown that the pseudogap state in the moir\'e triangular Hubbard model, in the presence of geometrically frustrated short-range antiferromagnetic correlations, is phenomenologically richer compared to the case of a square lattice and cuprates. It is not necessarily accompanied with an energy distribution curve (EDC) peak in the nodal direction and represents a more subtle state, which depends strongly on the short-range magnetic order and tight-binding structure, both of which can be further tuned by an additional parameter $\phi$ (corresponding to the degree of inversion symmetry breaking). Using CPT and comparing its results with GMFA, the evolution of the Fermi surface and the pseudogap in the triangular-lattice Hubbard model with the parameters $\phi$ and $U$ is explained from the viewpoint of the short-range magnetic order. We vary $\phi$ from 0 to $\pi/3$, in the range where the presence of antiferromagnetic correlations is expected \cite{Zang21, Wietek22}. The prominent signatures of the pseudogap are observed at low values of $\phi$: in this case, the direction $\Gamma-M$ (see Fig.~\ref{fig:Sketch}) has the most significant EDC drop at the Fermi surface. Also, at high values of $\phi$, there are signatures of the pseudogap with a drop in EDC in the $K-K'$ directions. Near $\phi = \pi/6$, the Fermi surface is sharp up to high interactions. In the wide range of $\phi$ and $U$, the signatures of Fermi arcs similar to those of cuprates are found. The types of Fermi arcs obtained within CPT are consistent with a simple GMFA picture, and, by comparing our CPT and GMFA results, the influence of short-range magnetic order on formation of the Fermi arcs is shown.

The rest of this paper is organized as follows. In Sec.~\ref{sec:2}, we briefly discuss the model and the methods. Section~\ref{sec:3} is devoted to the presentation of the CPT results. In Sec.~\ref{sec:4}, we discuss the CPT results and compare them with the GMFA results. In Sec.~\ref{sec:5}, the concluding remarks are given. The details concerning the calculations using a familiar example of a square lattice can be found in Appendix~\ref{sec:a}.

\section{\label{sec:2} Model and methods}
\subsection{Moir\'e Hubbard model}
\begin{figure}
\includegraphics[width=.8\linewidth]{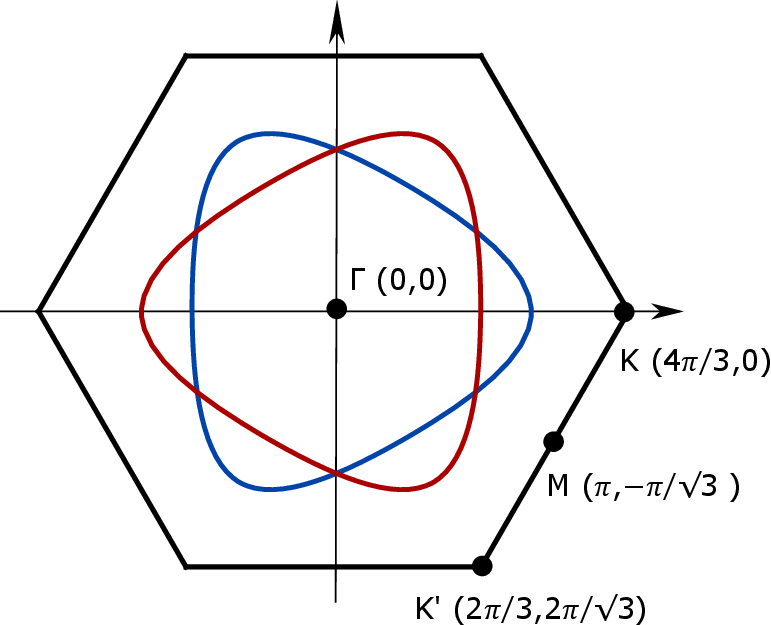}
\caption{\label{fig:Sketch} The sketch of the tight-binding Fermi surface in the first Brillouin zone at $\phi\sim\pi/12$. The blue (red) line illustrates the spin-up (spin-down) Fermi-surface.}
\end{figure}

 We study the moir\'e Hubbard model on a triangular lattice \cite{Wu18}
 \begin {equation}
H=-\sum\limits_{ij,\sigma} t_{i,j,\sigma} c_{i,\sigma }^{\dag} c_{j,\sigma}^{} + \sum\limits_{i}{U n_{i,\uparrow }{n_{i,\downarrow}}},
\end {equation}
where $c_{i,\sigma }$ denotes the annihilation operator of an electron on a site $i$ with spin projection $\sigma$, the particle number operator is $n_{i,\sigma }  = c_{i,\sigma }^\dag  c_{i,\sigma }^{}$, and $U$ is the on-site Coulomb interaction. Spin-dependent hopping is defined by the phase $\phi$, which determines the degree of the inversion symmetry breaking, as $t_{i,j,\sigma }  = \left| t \right|e^{i\sigma \eta _{i,j} \phi }$, where $\eta _{i,j}  =  \pm 1$, depending on the path connecting the nearest sites. Due to the Ising-like spin-orbit coupling, when $\phi \neq 0$, the electronic structure is nondegenerate with respect to spin, as schematically illustrated in Fig.~\ref{fig:Sketch}. The range of $\phi$ from $0$ to $\pi/3$ is studied, which is believed to be relevant for describing physically achievable values of displacement fields \cite{Zang21}. Since the paramagnetic phase is considered, the spectral function is shown only for spin-up electrons. The case of small finite doping $p=1/13$ is considered. The influence of hoppings beyond nearest neighbors is not considered in order to deal with basic Hubbard model physics only, not introducing another parametric dependence.

\subsection{Cluster perturbation theory}
In this paper, the electronic structure of the moir\'e Hubbard model is studied mainly within CPT \cite{Senechal00, Senechal02}. We study the electronic spectral function $A\left(\mathbf{k},\omega \right) = -{\dfrac{1}{\pi} \text{Im} G\left(\mathbf{k},\omega\right)}$, where $G\left(\mathbf{k},\omega\right)$ is the spin-up electron Green function, $\mathbf{k}$ is a wave vector and $\omega$ denotes frequency, which, along with other energy quantites, is measured in the units of the hopping integral $|t|$. CPT allows calculation of the electronic spectral function while explicitly taking into account correlations within a small cluster and is characterized by a fast convergence of results with increasing cluster size \cite{Huang22}. Intracluster interactions are taken into account by means of exact diagonalization of a cluster with open boundary conditions with the aid of the block Lanczos method \cite{Seki18}, while intercluster hoppings are treated perturbatively. The calculations are performed for the case of finite temperature $T=0.15$ to smooth out the spectra and thus reduce the influence of finite-size artifacts. However, since we study the paramagnetic non-superconducting phase, it does not affect the results significantly.

\begin{figure}
\includegraphics[width=.8\linewidth]{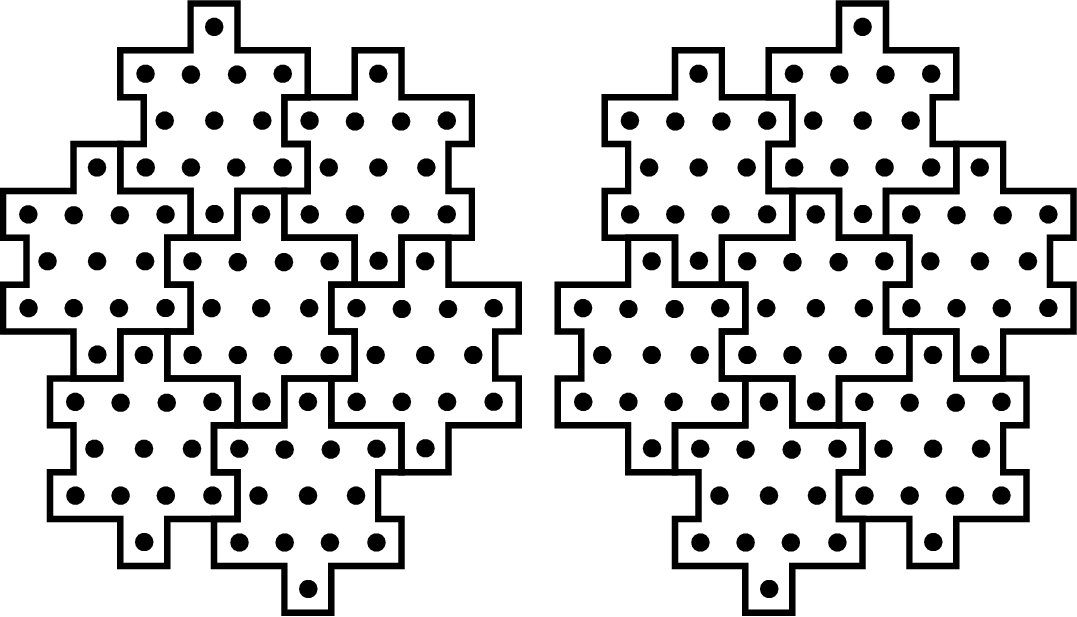}
\caption{\label{fig:Lattice} The two tilings used to obtain the hopping matrix with a 13-site cluster.}
\end{figure}

The main CPT equation is
\begin {equation}
\hat{G}\left(\mathbf{\tilde{k}},\omega\right)^{-1} = \hat{G}^c\left(\omega\right)^{-1} -\hat{T}\left(\mathbf{\tilde{k}}\right),
\end {equation}
where the quantities included in the equation are matrices with indices running through the intracluster sites, $\mathbf{\tilde{k}}$ is a wave vector defined within the cluster Brillouin zone, $\omega$ is frequency, $\hat{G}^c\left(\omega\right)$ is the cluster exact Green function, and $\hat{T}\left(\mathbf{\tilde{k}}\right)$ is a Fourier transform of the hopping matrix. As a final step, translational invariance is restored as
\begin {equation}
G\left(\mathbf{k},\omega\right) = \sum_{i, j} e^{i\left(\mathbf{r}_i - \mathbf{r}_j\right)\mathbf{k}} {G_{i, j}\left(\mathbf{k},\omega\right)},
\label {eq:CPT_translation}
\end {equation}
where $i(j)$ is an intracluster site index and $\mathbf{r}_{i(j)}$ is a corresponding radius-vector.
The lattice is covered by translations of a 13-site cluster in two ways, as shown in Fig.~\ref{fig:Lattice} with averaging $\hat{T}\left(\mathbf{\tilde{k}}\right)$ over the two tilings.

\subsection{GMFA}

It is convenient to have not only numerical data but also some analytical expressions for further analysis. In this paper, we refer to the GMFA as the simplest version of the two-time temperature Green function equations of motion method \cite{Zubarev60} combined with the Mori projection technique \cite{Mori65}, when the irreducible part of the multiparticle Green function entering the equations is neglected. We further supply the equations with a couple of fitting parameters in order to compare with CPT (see Sec.~\ref{sec:4}). For more elaborate applications of the technique, see Refs.~\onlinecite{Plakida13,Korshunov07,Shneyder20}.

Using the basis of one-site Hubbard operators $X^{p,q} = \left|p\right\rangle\left\langle q\right|$, the spin-up electron creation operator can be written as
\begin {equation}
a^{\dag}_{i,\uparrow} = X^{\uparrow, 0}_{i} + X^{2, \downarrow}_{i}.
\end {equation}
The normal part of the Green function defined as the matrix
\begin {equation}
D_{ij} = \left\langle {\left\langle {\left. {\hat \Psi _i } \right|\hat \Psi _j^\dag} \right\rangle } \right\rangle,
\end {equation}
where 
\begin {equation}
\hat \Psi _j^\dag = \left(X^{\uparrow, 0}_{j}, X^{2, \downarrow}_{j} \right),
\end {equation}
is given by the expression 
\begin {equation}
\hat D = \hat Q \left( {\omega \hat I - \hat E } \right)^{ - 1},
\end {equation}
where $\hat I$ is the unity matrix, $\hat Q = {\left\langle {\left. {\hat \Psi _i } \right|\hat \Psi _i^\dag} \right\rangle }$ (the space index is omitted in the paramagnetic phase) is equal to

\begin {equation}
\hat Q = \left( {\begin{array}{*{20}c}
   {Q_1 } & 0  \\
   0 & {Q_2 }  \\
\end{array}} \right) = \left( {\begin{array}{*{20}c}
   {1 + p} & 0  \\
   0 & {1 - p}  \\
\end{array}} \right)/2,
\end {equation}
where $p$ stands for doping concentration and the spectrum is determined from the linearization condition
\begin {equation}
\left\langle \left[ \left[\hat \Psi _i , H\right] - \sum\limits_{l} E_{i,l}\hat \Psi _l,\hat \Psi _j^\dag \right]_+ \right\rangle = 0.
\end {equation}
By calculating the averages of anticommutators entering the last equation in the simplest approximation when $\left\langle n_i n_j \right\rangle \approx \left\langle n_i \right\rangle \left\langle n_j \right\rangle$ and neglecting one-particle nonlocal correlations, which does not cause a significant effect on the results in the considered range of parameters, and then transforming to momentum space, we obtain the following expression for the Green function:
\begin {equation}
\hat D\left(\mathbf{k},\omega\right)  = \hat Q\left[ {\omega \hat I - \hat E\left(\mathbf{k}\right) } \right]^{ - 1}.
\end {equation}
Here, the spectrum is defined as
\begin {eqnarray}
\begin{array}{l}
 \hat E = \left( {\begin{array}{*{20}c}
   {\varepsilon _{11} \left( {\mathbf{k}\,} \right)} & {\varepsilon _{12} \left( {\mathbf{k}\,} \right)}  \\
   {\varepsilon _{21} \left( {\mathbf{k}\,} \right)} & {\varepsilon _{22} \left( {\mathbf{k}\,} \right)}  \\
\end{array}} \right), \\ 
 \varepsilon _{11} \left( {\mathbf{k}\,} \right) = Q_1^2 t\left( {\mathbf{k}} \right) + \alpha _c \frac{3}{N}\sum\limits_q {t\left( {\mathbf{k}\, - \,\mathbf{q}} \right)S\left( \mathbf{q} \right)}  \\ 
 \varepsilon _{12} \left( {\mathbf{k}\,} \right) = \varepsilon _{21} \left( {\mathbf{k}\,} \right) = Q_1 Q_2 t\left( {\mathbf{k}} \right) - \alpha _c \frac{3}{N}\sum\limits_{\mathbf{q}} {t\left( {\mathbf{k}\, - \,\mathbf{q}} \right)S\left( \mathbf{q} \right)}  \\ 
 \varepsilon _{22} \left( {\mathbf{k}\,} \right) = Q_2 U + Q_2^2 t\left( {\mathbf{k}} \right) + \alpha _c \frac{3}{N}\sum\limits_{\mathbf{q}} {t\left( {\mathbf{k}\, - \,\mathbf{q}} \right)S\left( \mathbf{q} \right)} , \\ 
 \end{array}
\end {eqnarray}
where $t\left( {\mathbf{k}} \right)$ is the Fourier transform of the spin-up hopping integral and $S\left( \mathbf{q} \right)$ is the Fourier transform of static spin correlations. Note that the parameter $\alpha _c$ is introduced in order to analyze the role of spin correlations in the discussion below. The physical meaning of the phenomenological parameter $\alpha_c$ is that it regulates the sensitivity of the electronic subsystem to the influence of spin correlations. In the simplest mean-field approximation known as the Hubbard-I, $\alpha_c$ is equal to zero. In our GMFA we fit this parameter to reproduce the CPT data. We obtain an estimate of $S\left( \mathbf{q} \right)$ from the exact diagonalization of a cluster for the appropriate parameters of the model by restoring translational invariance in a CPT-like manner:
\begin{equation}
S\left( \mathbf{q} \right) = \sum\limits_{i,j} {e^{i\left( {{\mathbf{r}}_i  - {\mathbf{r}}_j } \right)\mathbf{q}} } \left\langle {S_i^z S_j^z } \right\rangle, 
\label{eq:S_q}
\end{equation}
where the sum runs over the intracluster indices.

\section{\label{sec:3} Electronic structure and spin correlation functions within CPT}

\begin{figure}
\includegraphics[width=1.0\linewidth]{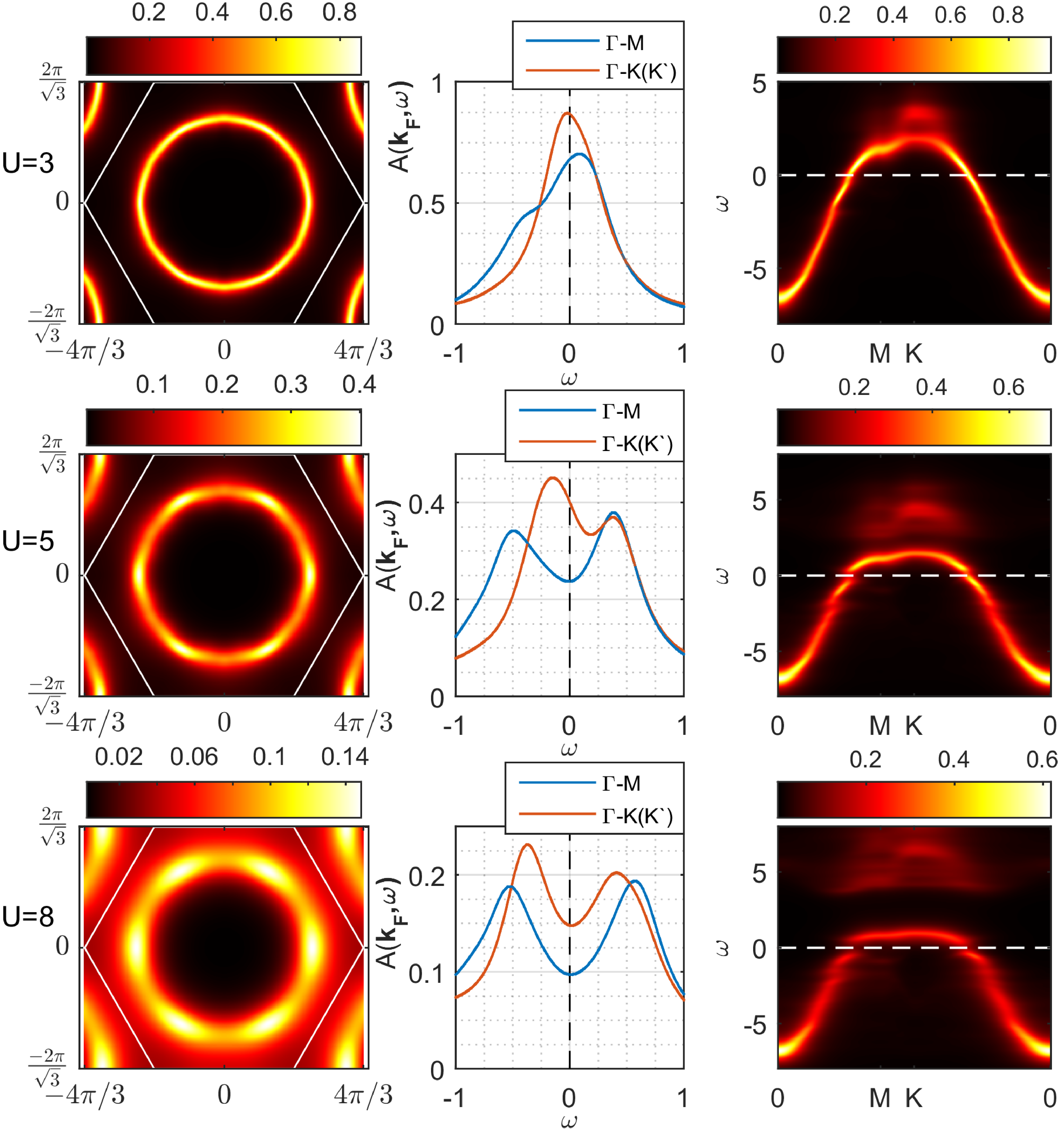}
\caption{\label{fig:Phi0} The electronic structure of the model with $\phi=0$ obtained with CPT for the three values of the Coulomb repulsion shown for one spin projection. Left column shows the Fermi surface evolution with interaction. Middle column shows the frequency-dependent spectral functions $A\left(\omega\right)$ for the $k$-points with maximal weight at the Fermi surface in the directions shown in the insets. Right column presents the spectral function high-symmetry cuts, the momentum dependence of the intensity peak demonstrates the quasiparticle dispersion with variable spectral intensity along the dispersion curves.} 
\end{figure}

\begin{figure}
\includegraphics[width=1.0\linewidth]{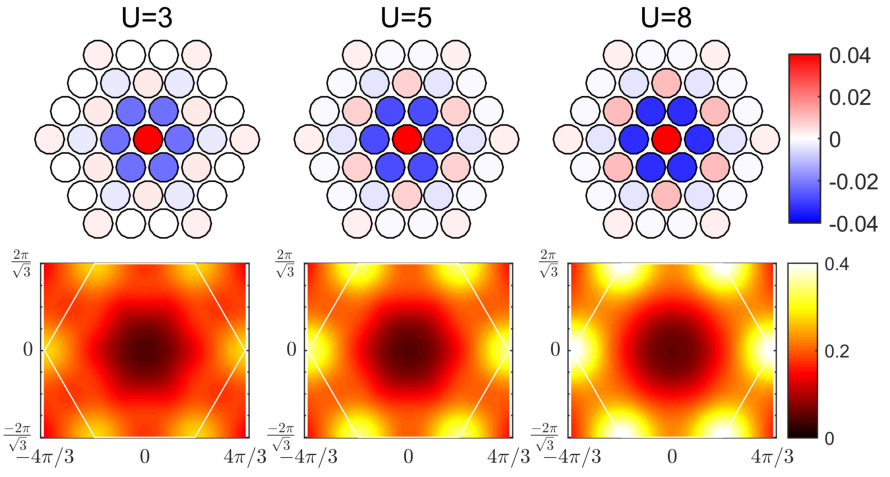}
\caption{\label{fig:S_phi0} The real-space spin-correlations (upper row) $S\left(\mathbf{r}\right)$ with $\mathbf{r}=0$ being the central red site and the spin structure factor $S\left(\mathbf{q}\right)$ (lower row) obtained using CPT-like scheme for restoring translational invariance (Eq. \ref{eq:S_q}).}
\end{figure}

First, we study the case of the standard Hubbard model when $\phi=0$ and the momentum points $K$ and $K'$ are equivalent. The evolution of the Fermi surface with Coulomb interaction is shown in Fig.~\ref{fig:Phi0}. With increasing interaction, the circle-like Fermi surface at small interaction $U=3$ when the Hubbard bands are almost merged (see the right column of the figure), acquires six arc-like regions located in the $\Gamma-K$-directions at intermediate interaction $U=5$ (when the Hubbard bands start to split) and strong interaction $U=8$ (when there is a clear splitting between the lower and upper Hubbard bands). In what follows, we call ``the spectral function $A\left(\omega\right)$ in the direction $X-Y$'' a quantity defined as $A\left(\mathbf{k}_0, \omega\right)$, where $\mathbf{k}_0$ is a point with maximum spectral weight at the Fermi level along the direction $X-Y$. Already at rather small $U=3$ there can be seen some non-equivalency between the frequency-dependent spectral functions in the $\Gamma-M$ and $\Gamma-K$ directions: the spectral peak for $\Gamma-M$ is lower. At $U=5$, the $\omega$-dependent spectral weight curve is reminiscent of the pseudogap in cuprates: while the "nodal" $\Gamma-K$ direction is characterized almost by a peak at the Fermi level, the "antinodal" $\Gamma-M$ direction has a dip at $\omega=0$. Finally, at $U=8$, there is a momentum-dependent suppression at $\omega=0$ in both directions, with the largest spectral weight in the $\Gamma-K$ direction. As shown in Fig.~\ref{fig:S_phi0}, the changes in the electronic structure are accompanied by the growth of antiferromagnetic short-range order. The considered situation differs from the square-lattice cuprate-like case, with weak geometrical frustration and stronger antiferromagnetic correlations, where a strong peak is usually seen in the nodal direction.

\begin{figure*}
\includegraphics[width=1.0\linewidth]{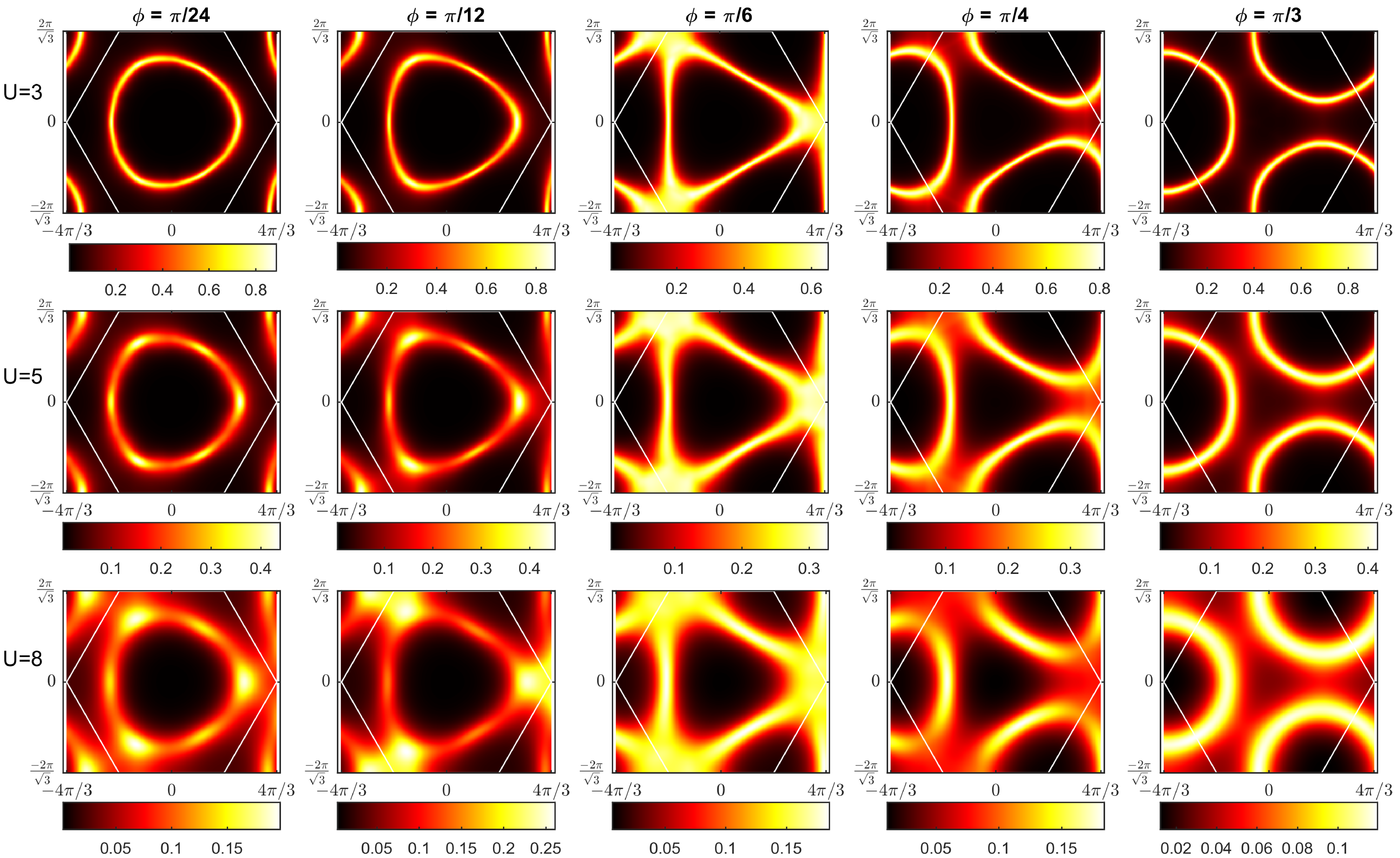}
\caption{\label{fig:Fermi} The spin-up Fermi surface evolution (within each column) with interaction obtained using CPT for several values of parameter $\phi$.}
\end{figure*}

The changes in the Fermi surface with $\phi$ and $U$ are shown in Fig.~\ref{fig:Fermi} for the spin-up projection (the spin-down is related to it by a $\pi$-angle rotation). For a relatively small $\phi = \pi /24$, at small $U$, the Fermi surface is characterized with an almost uniform spectral weight distribution, but increasing $U$ leads to the manifestation of arc-like areas in the directions $\Gamma-K$ and $\Gamma-K'$ with suppressed spectral weight in the $\Gamma-M$ direction, as in the case of zero $\phi$. However, the arcs in the $\Gamma-K$ directions are clearly more pronounced than in $\Gamma-K'$, which means the spin-polarized momentum-selective behavior. A similar tendency is observed at $\phi = \pi / 12$, but suppression of the Fermi-level peak for the $\Gamma-K$ direction is not observed, likely because the Fermi level becomes closer to the Van-Hove-like singularity. At $\phi$ near $\pi/6$ the Fermi level is close to the Van-Hove-like singularity at small $U$, where the triangular Fermi surface is the transition point to a new topology. At $\pi/6$, with increasing U, the arc-like distribution of spectral weight manifests itself, now in a $\Gamma-K'$ direction. At higher $\phi$ values $\phi = \pi/4, \pi/3$ the Fermi surface at weak interaction is located around $K'$ points, as opposed to the $\Gamma$ point at low $\phi$. For $\phi=\pi/4$, increasing the interaction leads to arc-like areas located pronounced in the $\Gamma-K'$ directions. At $\phi = \pi/3$, no significant momentum-space anisotropy is detected.

\begin{figure*}
\includegraphics[width=1.0\linewidth]{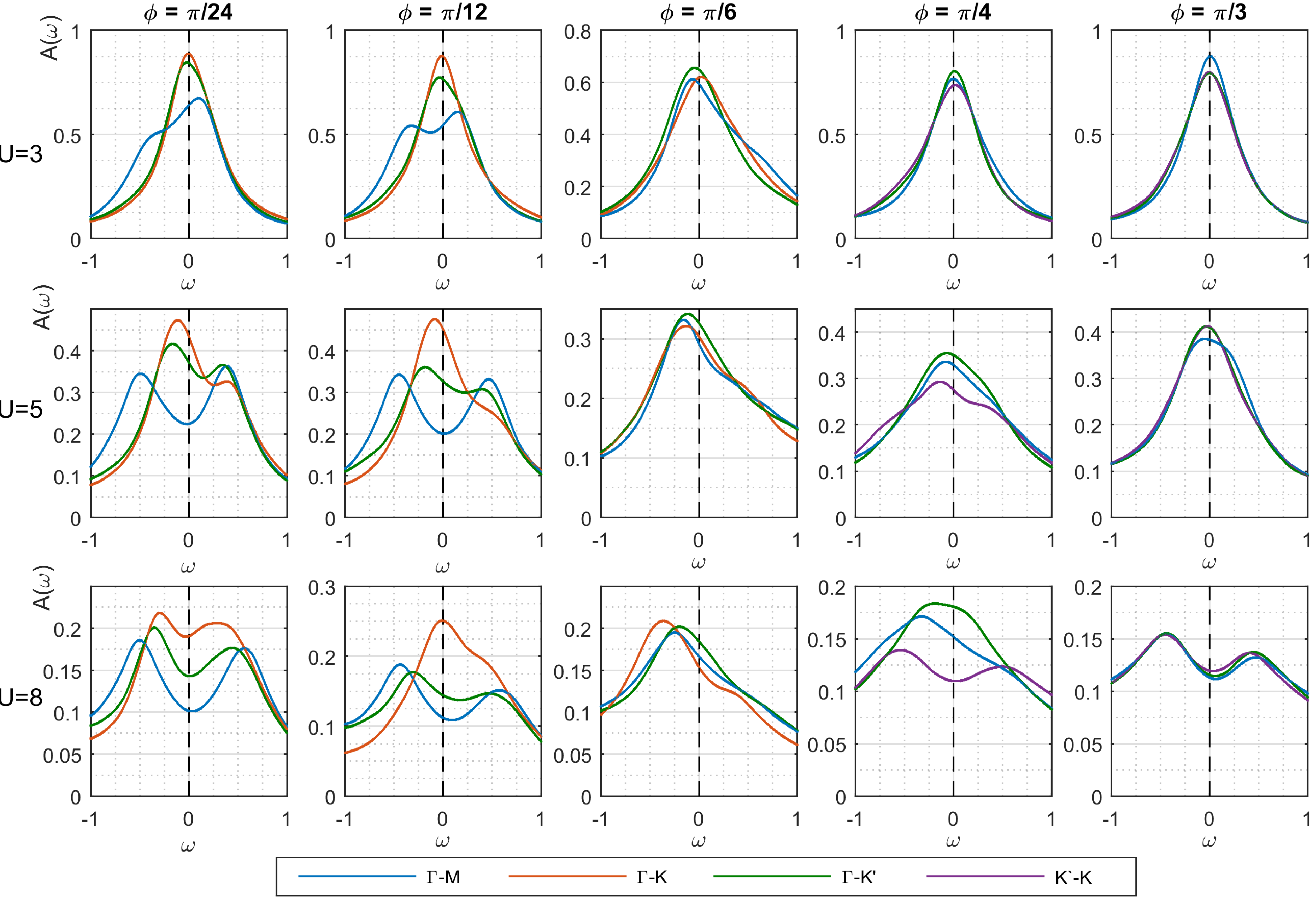}
\caption{\label{fig:Aw} The evolution with interaction of the spin-up frequency-dependent spectral function obtained with CPT for the $k$-points with maximal weight at the Fermi surface in the directions shown in the inset for several values of parameter $\phi$.}
\end{figure*}

In order to demonstrate the frequency-dependent behavior, in Fig.~\ref{fig:Aw} we show the spin-up spectral function $A\left(\omega\right)$ along some special directions. For the smallest spin-orbit coupling considered here, at $\phi=\pi/24$, the overall picture is similar to the case of $\phi=0$, the main difference is that the spectral weight of the arcs in $\Gamma-K$ direction is larger than in $\Gamma-K'$. At $U=3$ there is already some anisotropy, resulting in the suppressed spectral peak for the $\Gamma-M$ direction. When the Coulomb interaction increases from $U=3$ to $U=5$, the behavior changes qualitatively as the dip at the Fermi level appears for $\Gamma-M$. At $U=8$ the suppression is present for all directions, but is stronger for $\Gamma-M$. The spectral weight distribution at larger $\phi=\pi/12$ is similar, but at $U=8$ there is a peak at the Fermi level for the $\Gamma-K$ direction, which is due to the underlying tight-binding dispersion that stabilizes the spectral weight near the $K$ point. At $\phi=\pi/6$, the system is close to the Van-Hove-like singularity, and the spectral peaks are well defined for all directions. However, at higher interactions, there is some degree of anisotropy at the Fermi surface and a tendency to arc formation. At $\phi=\pi/4$, due to the change in the Fermi surface topology, the most pronounced peak is in the $\Gamma-K'$ direction, and in this case a dip with increasing interaction is formed for the momentum direction along the boundary of the Brillouin zone, $K'-K$. For the largest spin-orbit coupling considered here, $\phi=\pi/3$, no significant momentum space anisotropy is considered, but at $U=8$ a bad metal behavior with suppression of the spectral weight at the whole Fermi surface is seen.

\begin{figure}
\includegraphics[width=1.0\linewidth]{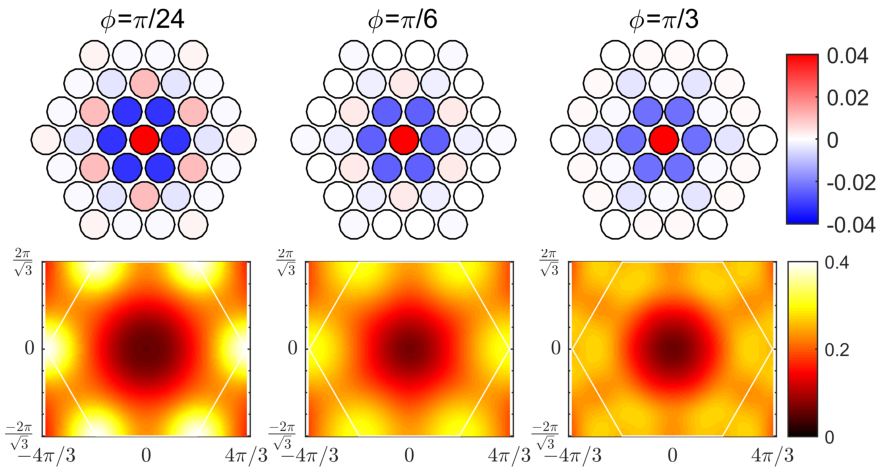}
\caption{\label{fig:S_U8} The real-space spin-correlations (upper row) $S\left(\mathbf{r}\right)$ with $\mathbf{r}=0$ being the central red site and the spin structure factor $S\left(\mathbf{q}\right)$ (lower row) obtained at $U=8$ using CPT-like scheme for restoring translational invariance (Eq. \ref{eq:S_q}).}
\end{figure}

As shown in Fig.~\ref{fig:S_U8}, increasing spin-orbit interaction in this case leads to an overall tendency to decrease short-range ordering. From the upper panel of the figure it is seen that the second neighbor (ferromagnetic) correlations decrease, while the first (antiferromagnetic) weaken and the third (antiferromagnetic at small $\phi$) remain almost unchanged. The maxima of the structure factor also weaken, and, at $\phi=\pi/3$, the maximum of the structure factor is shifted from the $K$ point. These changes in short-range magnetism correlate with the decreasing of the tendency towards formation of the Fermi arcs, which is larger in the case of pronounced antiferromagnetic correlations. However, even at $\phi=\pi/3$, the correlation effects are significant at high values of Coulomb repulsion, leading to a pseudogap-like suppression of the density of states, but uniformly in the momentum space, which is not characteristic of the most familiar kind of the pseudogap in the presence of strong antiferromagnetic correlations.

\section{\label{sec:4} Discussion}
When using the exact diagonalization solver, different physical mechanisms are strongly intertwined, and it is difficult to tell which affects the results the most. Finite-range spin correlations have been shown to be crucial for the pseudogap \cite{Senechal04,Stanescu06,Ye23}. 
Within CPT itself, it is hard to identify what is the cause of the pseudogap-like effects in the results presented above. Moreover, we found that the effects of cluster size and shape in CPT are severe when studying the electronic structure of the Hubbard model on the triangular lattice using small clusters. Thus, it is useful to verify whether, at least qualitatively, the class of Fermi surfaces obtained in CPT agrees with the results obtained using a method that does not break translation invariance and contains adequate physical mechanisms. 
In this way, to compare with CPT, we apply a semi-analytical approach of GMFA, supplied with a phenomenological fitting parameter $\alpha_c$, which regulates the strength of coupling between the electronic structure and spin correlations, as described in the section~\ref{sec:2}. The method of GMFA in the simplest framework applied in this paper contains two contributions. The first is the Hubbard-I dispersion dependent on the hopping integrals and Coulomb repulsion. The second is a correction due to spin correlations. In order to compensate for the effects beyond the lowest-order perturbation theory, which are present in CPT and absent in GMFA (see Appendix~\ref{sec:a}), we add the effective hopping integrals between the second and third neighbors to our GMFA calculations. Here, these tight-binding fitting parameters are: $t'_{eff}=0.1, t''_{eff}=-0.05$. Note that we study the model with only nearest hoppings with our main method, i.e. CPT.
The effective hoppings and $\alpha_c$ are chosen in such a way that they reproduce qualitatively the CPT spectral weight distribution at the Fermi level when $\phi=0$. These parameters are fixed when varying $\phi$. We draw the reader's attention that the version of GMFA used in this paper is too simple to allow for the correct description of the pseudogap drop in the EDC: within GMFA, the pseudogap suppression of spectral weight can appear technically either from the irreducible part of the equations of motion or from additional quasiparticles introduced, which is not taken into account in the present work. However, it turns out that the Fermi surface evolution in cuprates within this method is adequately described at the quantative level \cite{Shneyder12}. Thus, we compare only Fermi surfaces. In Appendix~\ref{sec:a} we discuss the comparison of GMFA and CPT Fermi surfaces on a well-known example of a square lattice to illustrate the influence of the effective hoppings and $\alpha_c$.

\begin{figure*}
\includegraphics[width=1.0\linewidth]{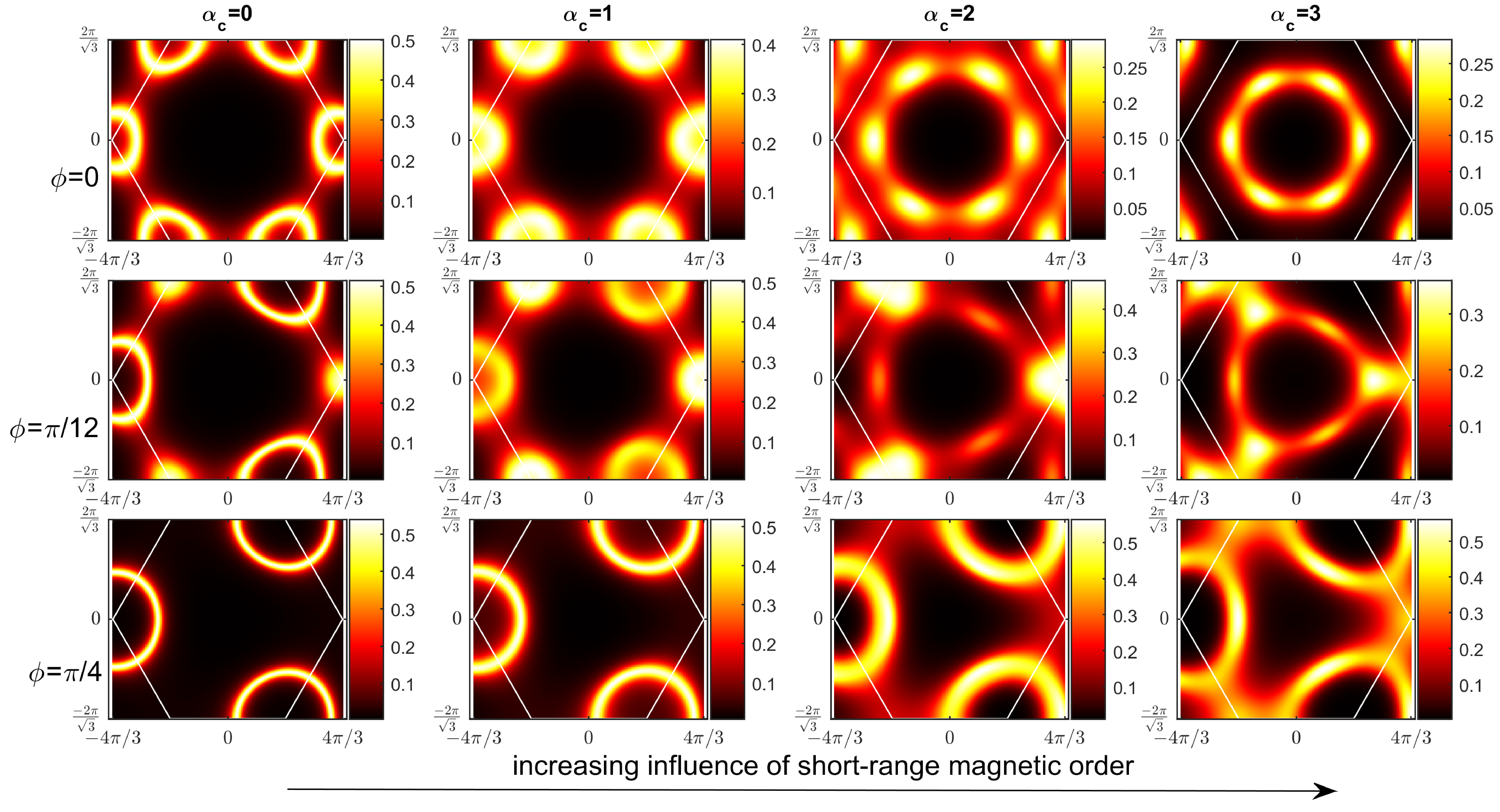}
\caption{\label{fig:GMFA} The dependence on the parameter $\alpha_c$ in the GMFA results for $U=8$ and $\alpha_c$ = \{0,1,2,3\}. The Lorentzian broadening is the same as in the CPT calculations, $\delta = 0.24$.}
\end{figure*}

In Fig.~\ref{fig:GMFA} we present the dependence of the Fermi surface on the parameter $\alpha_c$ for several values of $\phi$. At $\alpha_c=0$, which corresponds to the Hubbard-I approximation, at $\phi=0$ the Fermi surface consists of six pockets with uniform distribution of spectral weight around the $K(K')$ points. As we increase $\alpha_c$, the Fermi surface undergoes a transition to six arc-shaped areas concentrated around the center of the Brillouin zone. At moderate $\phi$, as shown in the second row of Fig.~\ref{fig:GMFA}, at the Hubbard-I level, the pockets around the points $K$ and $K'$ are “small” and “big”, respectively. As the influence of two-particle correlations increases, the pockets transform into the arc-like segments with the “small” ("big") one has a tendency to have larger (smaller) spectral weight. At quite large $\phi$, as the third row of Fig.~\ref{fig:GMFA} shows, when the spin-spin correlations are weaker and lose antiferromagnetic character, no such radical change when turning on $\alpha_c$ as in previous cases is detected: the main effect is in arc-like distribution of spectral weight along the Fermi surface around the $K'$ points starts to form. Including static spin-spin correlations in GMFA produces Fermi surfaces in agreement with those obtained within CPT (see the right half of Fig.~\ref{fig:GMFA}). This gives evidence in support of the consistence of the CPT results with a simple physical mechanism: the short-range magnetic order together with the dispersion, which is split into two zones by the Coulomb repulsion, play major role in the formation of the observed spectral properties.

It should be noted that CPT has a tendency to overestimate the influence of correlation effects. Compared to DCA studies \cite{Chen13,Downey23,Downey24}, the pseudogap in CPT is observed at lower interactions and higher doping values. However, within the pseudogap phase, the spectral weight distribution obtained using interpolation within DCA at half filling \cite{Chen13} is similar to the arc-like structures obtained here at small doping. The main spectral weight contribution at the Fermi surface is consistent with the CPT and its variational extension study \cite{Zong24} at half filling and reasonably small $\phi$ (note that at large $\phi$ the parameters in that paper account for a strong influence of fourth neighbors as opposed to the nearest-neighbor case here). However, using a 13-site cluster in our case does not lead to splitting of the Fermi surface in segments, as opposed to the 12-site case. In the well-known case of a square lattice, the pseudogap state has been shown to depend strongly on short-range correlations \cite{Senechal04, Stanescu06}, and 12- and 16-site clusters are commonly used in CPT. It is important to have the characteristic spin correlation length within the cluster. From this point of view, a 12 or 13-site cluster in a frustrated triangular lattice system with a reduced correlation length (compared to the square lattice case) should be of sufficient size. However, shape effects can be significant, poorly controlled, and can lead to the appearance of artificial density waves \cite{Verret19} affecting the details of the spectral weight distribution. Thus, in the case of a triangular model studied here, further studies of the momentum dependence of the spectral function are needed to confirm the Fermi surfaces presented here using methods capable of taking into account larger structures, such as the time-dependent density matrix renormalization group methods \cite{Yang16} or quantum Monte Carlo \cite{Huang22}. We expect that the main effects, such as the influence of short-range magnetic order on the pseudogap, should be preserved when using larger lattices, but it would be interesting to improve the details regarding the $\mathbf{k}$ dependence and the EDC curves for the specific triangular model. The main test of the Fermi surfaces obtained here is their agreement with the known half-filling results from the literature in general, as well as with the mean-field calculations using GMFA, which have a very clear physical meaning. 

\section{\label{sec:5} Conclusion}

In this paper, the evolution of the Fermi surface and the pseudogap with parameters, which can arguably be varied when simulating the Hubbard model experimentally in moir\'e systems, has been studied. The inversion symmetry breaking parameter $\phi$ and Coulomb repulsion $U$ were varied within the region where the magnetic correlations change from antiferromagnetic to those with a tendency towards ferromagnetism. The results were obtained using CPT in the paramagnetic phase of the lightly doped triangular-lattice moir\'e Hubbard model and explained from the viewpoint of short-range magnetic order with the aid of mean-field calculations. It was shown that, in agreement with the short-range spin correlations behavior, the pseudogap is observed at low values of $\phi$, with the direction $\Gamma-M$ having the most prominent EDC pseudogap drop at the Fermi surface. Close to $\phi=\pi/6$ the Fermi surface is sharp up to high interactions due to the strong influence of the underlying tight binding structure, which exhibit a Van Hove singularity close to the Fermi level. At high values of $\phi$, there are signatures of the pseudogap in the $K-K'$ directions, but weaker than those obtained in the presence of a strong antiferromagnetic short-range order.

The results obtained in this paper highlight the possible role of nonlocal magnetic correlations in novel moir\'e systems such as $t$WSe$_2$. It is shown that the pseudogap behavior in the geometrically frustrated system supplied with an additional parameter has a more rich phenomenology compared to the well-known square-lattice cuprate type pseudogap. Here, the pseudogap is not necessarily accompanied by an EDC peak in the nodal direction: in our frustrated case, the momentum-dependent suppression of the spectral weight at the Fermi level for a particular spin projection can be observed for all directions. The experimental observation of the Fermi arc like structures and pseudogap features in moire systems such as obtained here, together with spin-polarized pseudogap behavior, is technically challenging due to the need to resolve momentum-space peculiarities within the small moire superlattice Brilluoin zone, but may be accessible in future studies using the state-of-art ARPES methods with fine momentum resolution \cite{Chen24}. 

\appendix
\section{\label{sec:a} Comparison of GMFA and CPT on a square lattice}
\begin{figure}
\includegraphics[width=1.0\linewidth]{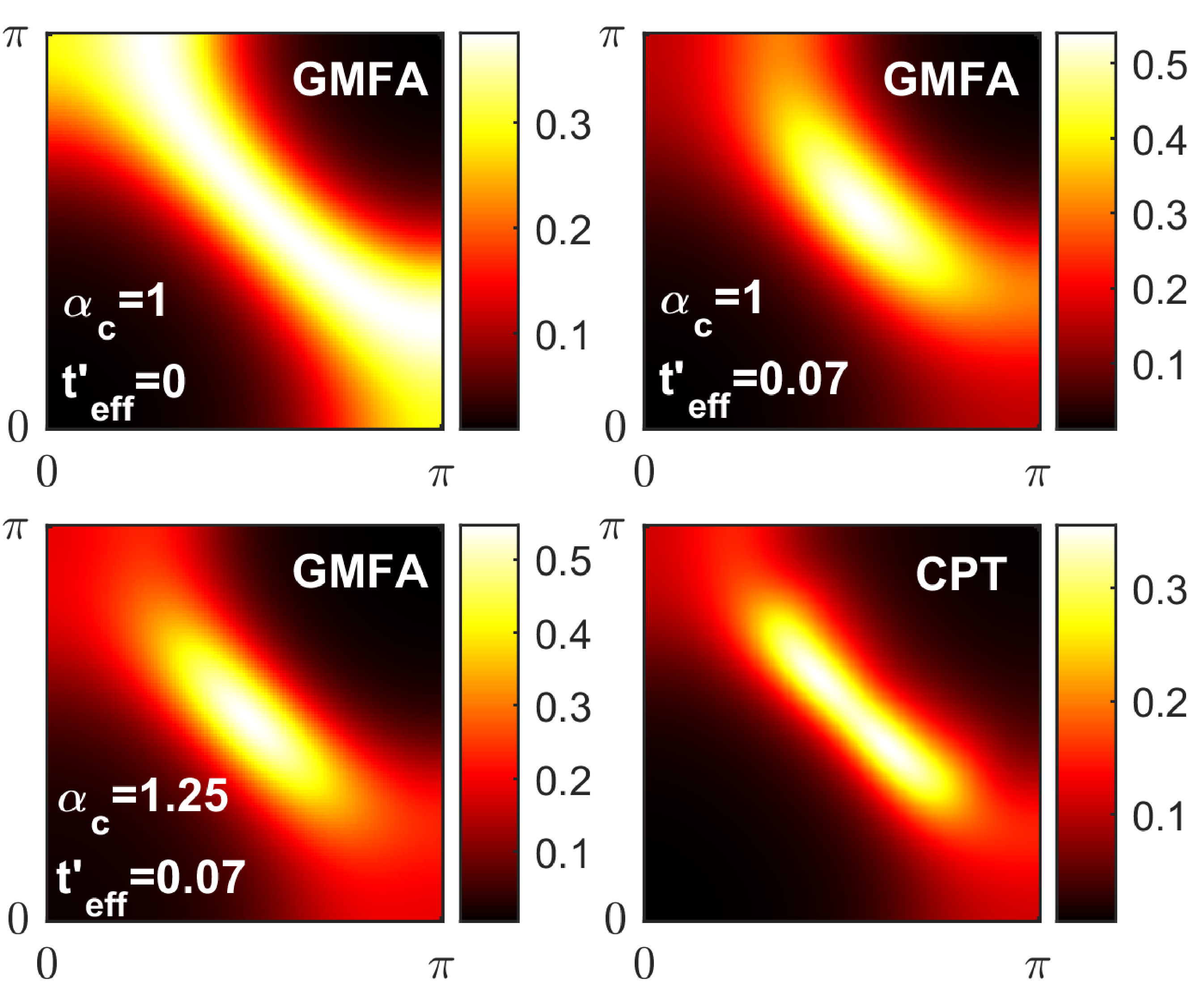}
\caption{\label{fig:GMFA_square} Comparison of the Fermi surface obtained within GMFA using various fitting parameters and using CPT for the Hubbard model on a square lattice at doping $p=1/16$ and $U=8$; the Lorentzian broadening is $\delta=0.16$.}
\end{figure}

The Fermi surface with uniform spectral weight distribution obtained within GMFA without additional fitting for the Hubbard model on a square lattice does not reproduce the well-known arc, which is seen in CPT, as illustrated in Fig.~\ref{fig:GMFA_square}. However, introducing effective hopping $t'_{eff}$ leads to the non-uniform spectral weight distribution, as shown for $t'_{eff}=0.07$ in Fig.~\ref{fig:GMFA_square}. It can be assumed that there is some degree of effective hopping of higher order in perturbation theory than the version of a mean-field approach presented here accounts for. Moreover, the increased coupling, $\alpha_c$, between the electronic structure and the spin correlations leads to an improved agreement between the CPT and GMFA results, most likely due to a rough compensation for the lack of correlation effects in the simplest mean-field approximation. This example demonstrates that simple account for static short-range correlations and effective hopping gives a phenomenological description of the Fermi-arc on a mean-field level.

\begin{acknowledgments}
Sections ~\ref{Intro},~\ref{sec:4} and~Appendix~\ref{sec:a} have been carried out along the state assignment of the Kirensky Institute of Physics, and ~\ref{sec:2}, ~\ref{sec:3}, ~\ref{sec:5} - with the support of the Russian Science Foundation, project no. 24-12-00044.
\end{acknowledgments}

\bibliography{paper}

\end{document}